\documentclass[doublecol]{epl2} 
\usepackage{amssymb}
\newcommand{\half}{\frac{1}{2}}
\newcommand{\thrd}{\frac{1}{3}}
\newcommand{\frth}{\frac{1}{4}}
\newcommand{\sxth}{\frac{1}{6}}

\newcommand{\tauv}{\mbox{\boldmath $\tau$}}
\newcommand{\tW}{\tauv{\bf \cdot W}}

\newcommand{\delv}{\mbox{\boldmath $\nabla$}}
\newcommand{\sigv}{\mbox{\boldmath $\sigma$}}
\newcommand{\Astr}{/\!\!\!\!A}
\newcommand{\pstr}{/\!\!\!p}
\newcommand{\kstr}{/\!\!\!k}

\newcommand{\lvac}{\langle0|}
\newcommand{\rvac}{|0\rangle}
\newcommand{\sqr}{\square}
\title{
Neutrino and scalar boson mass in algebraic quantum field theory }
\author{R. K. Nesbet }
\institute{ 
IBM Almaden Research Center,
650 Harry Road,
San Jose, CA 95120-6099, USA
}
\pacs{11.15.-q}{Gauge field theories}
\pacs{14.60.Pq}{Neutrino mass and mixing}
\pacs{14.80.Bn}{Standard-model Higgs bosons}
\abstract{
The hypothesis is explored that fermion rest mass is due entirely to
self-interaction via virtual excitation of gauge bosons.
This requires revising the standard model to treat both chiral
projections of a fermion field as SU(2) doublets, which precludes
Yukawa coupling to a scalar (Higgs) boson field.
The estimated self-interaction mass of the electron neutrino is
$0.291\times10^{-5}m_e$.  The implied self-interaction mass of the Higgs
boson itself is very small, comparable to the neutrino.  Because there
is no direct coupling to fermions, only to the $Z^0$ gauge boson, this
can be reconciled with failure to detect low-mass Higgs bosons.  This
argument eliminates many undetermined parameters of the standard model,
but requires an {\it ad hoc} Lagrangian term to account for neutral
current asymmetries.  The proposed algebraic formalism is consistent
with fermion generations defined by distinct eigenvalues of a
self-interaction mass operator. 
}
\begin{document}
\maketitle
\begin{center} For {\em Europhysics Letters} \end{center}
\section{The standard model revisited}
\par The standard model(SM)\cite{WEI96,CAG98}, based on conventional 
quantum field theory\cite{PAS95}, explains qualitatively and in many
cases quantitatively almost all established experimental data for
elementary particles and fields other than gravity. A striking exception
is neutrino mass, empirically small but nonzero\cite{NEU04}.  Although
nonvanishing neutrino mass appears to be empirically established,
the electroweak theory of Weinberg and Salam\cite{REN90,CAG98,WEI96}
constrains neutrino mass to vanish by excluding right-handed chiral
neutrino fields.  SM assumes left-chiral fermion projections to be SU(2)
doublets, while right-chiral projections are SU(2) singlets. 
Equivalently, chiral projectors modify gauge boson Feynman vertices.

\par The theory is assumed to descend by
successive symmetry-breaking from a renormalizable grand-unified
theory (GUT) of higher symmetry.  This excludes bare fermion mass, so
that bare fermion fields are chiral eigenstates.  Hence chiral
projection operators do not affect the parental GUT.  Their introduction
in SM justifies fermion mass generation by Yukawa coupling of fermion
fields to the SU(2) doublet Higgs scalar boson.  Such coupling terms 
in the Lagrangian are invariants only for fermion current densities
constructed from an SU(2) doublet field and a distinct SU(2) singlet
field\cite{GHKD90}.
\par Since neutrino mass requires some modification of the theory, it is
of interest to see whether the large number of arbitrary parameters 
required by SM might be reduced, without invoking unknown new physics or
departing from quantum field theory.  Both chiral projections of fermion
fields are assumed here to be SU(2) doublets, eliminatng inequivalent
field projections.  This assumption excludes Higgs-fermion Yukawa 
coupling terms from the Lagrangian, because they cannot be SU(2)
invariants.  In accord with the logical principle of `Occam's razor',
removing chiral projection operators eliminates arbitrary mass  
parameters.  This can be done without affecting the success of SM
electroweak theory for processes involving only leptons. The very small
mass of neutrinos makes chiral projection redundant.  The modified
theory can be reconciled with neutral current asymmetries by inserting 
into the Lagrangian density an {\it ad hoc} term that does not add 
arbitrary parameters.
\par Self-interaction mass results from virtual emission and
reabsorption of gauge field quanta\cite{SCH58,FEY49,WEI39,PAS95}.  
Depending on the same operators as spontaneous emission, these processes
cannot be eliminated without selective chiral projections.  Otherwise, 
it is inconsistent to postulate neutrino mass exactly zero\cite{NES03}. 
For lepton/neutrino processes with intermediate charged $W^\pm$ 
gauge bosons it will be shown here that chiral projectors can be omitted
without affecting the established V-A phenomenology.  This
justifies a calculation of the electron-neutrino self-interaction 
mass consistent with current empirical data.  
\par An interacting massless bare fermion is dressed by virtual gauge
fields to become a quasiparticle. Stability
of the dressed field requires diagonalization of a mass operator defined
by gauge field coupling.  The Lagrange multiplier required for such
secular stability is a spacetime invariant mass.  Mass diagonalization 
defines a canonical transformation of the vacuum state.  An appropriate 
algebraic formalism, compatible with perturbation theory but not
dependent upon it, is used here to reconsider relevant implications
of quantum field theory. 
\par  In relativistic
perturbation theory\cite{FEY49}, self-interaction mass is a sum over
momentum transfer, logarithmically divergent\cite{WEI39,PAS95} unless
the sum is somehow cut off.  This divergence indicates that the theory
is incomplete.  The 4-momentum cutoff required for convergent integrals
is not determined by perturbation theory, and may be extrinsic to
quantum field theory.  Otherwise, electron mass is determined by
electromagnetic interaction.  The ultimate self-contained theory must
include a cutoff mechanism that implies correct masses for all fermions.
\par Extending QED to nonabelian gauge symmetry\cite{YAM54}, 
electroweak theory\cite{REN90,CAG98,WEI96} incorporates neutrinos, 
quarks and SU(2) weak gauge fields.  The mass of the Maxwell field is
forced to zero by decoupling from the Higgs scalar boson field. 
SM postulates Yukawa coupling of fermions to the Higgs field,
introducing coupling parameters adjusted to physical fermion masses.
If masses are determined by self-interaction induced by coupling to the
gauge fields, parametrized Higgs coupling is not needed.  
Since neutrinos are coupled only to the weak
gauge fields, their self-interaction mass must be small.
\par A significant implication of this analysis is that for fermions
the self-interaction mass operator might very well have several
eigenvalues that correspond to discrete states pushed down below
successive overlapping continua.  This could explain the existence of
higher generations of fermions, such as heavy leptons and their
corresponding neutrinos.  
\par Induced self-interaction is considered here for a scalar
boson field. In fact, Lagrangian terms implied by SU(2) gauge covariance
can replace the parametrized Higgs self-interaction, while
retaining the essential structure of electroweak theory.
If the Higgs mass results solely from its coupling to the weak gauge
fields, it might be very small.  
\par Renormalizable field theory with no bare mass is covariant under a
local scaling transformation.  If gravitational theory is modified to
incorporate this same conformal symmetry\cite{MAN06}, it identifies the
imaginary mass term in the scalar field equation with the Ricci scalar,
a measure of global spacetime curvature.  The self-interaction mass
considered here is not inconsistent with this gravitational theory.  It
augments the extremely small but nonzero implied curvature mass by a
much larger dynamical self-interaction, still much smaller than the
currently anticipated multi-GeV Higgs mass\cite{GHKD90,HIG06,OPL03}.
\par The algebraic formalism is applied here to consider the mass of the
electron, of the electron-neutrino, and of the scalar (Higgs) boson.
This analysis reproduces the Feynman electron self-energy at the lowest
level of approximation.  It justifies a
calculation that results in a neutrino mass consistent with current
empirical limits.  If similar logic is valid for the scalar boson,
its mass would be comparable to that of the neutrino.  This conflicts
with current expectations, but might merit reexamination of empirical
data if a multi-GeV Higgs boson is not found\cite{GHKD90}.  A neutral 
boson of very small mass, not coupled to fermions, might not be directly
observable.  A $Z^0$-induced Higgs cascade might be analogous to 
bremsstrahlung.

\section{Postulates of the theory}
\par That a physical fermion field is a quasiparticle which acquires
mass from its self-interaction is consistent with the structure of the
Dirac equation: a mass parameter couples field components of opposite
chirality whose energy values have opposite sign.  The field equation
for a bare fermion coupled to gauge fields can be rewritten so that its
mass is an eigenvalue of a scalar self-interaction mass operator.
Diagonalization in the Fock space of the interacting system defines a
canonical transformation from bare fermions to dressed quasiparticles,
identified as physical fermions. 
An algebraic formulation is developed here in which this
transformation, which breaks chiral symmetry as it produces 
nonvanishing mass, is explicit.  
\par The constraint of fixed normalization for timelike displacements
determines quasiparticle mass as an invariant Lagrange multiplier.  If a
quasiparticle state is embedded in a continuum of virtual excitations,
the implied entity is a resonance, with an inherent finite lifetime.
The secular stability of a quasiparticle constructed by diagonalization
is global in spacetime.  Matrix elements relevant to decay into
constituent fields are removed by construction.
This means in particular that a dressed electron,
built from chiral bare fields of positive and negative energy, cannot
decompose by interactions, short of a canonical transformation of the
renormalized vacuum.  Thus one cannot consider massive chiral fermions
to be elementary fields.  
\par Quantum field theory is based on two distinct postulates.
The first is the dynamical postulate that action integral
$W=\int{\cal L}d^4x$, defined by Lagrangian density ${\cal L}$
over a specified space-time region, is stationary with respect to
variations of the independent fields, subject to homogeneous boundary
conditions.  The second postulate,
which requires the fields to be operators in a Fock space, modifies and
constrains classical field theory.  Field operators act on state vectors
defined by virtual excitations of a vacuum state.  Field equations for
interacting field operators imply coupled algebraic equations when
projected onto the state vectors.
\par For Dirac field $\psi$ and Maxwell field $A_\mu$, defining
$F_{\mu\nu}=\partial _{\mu} A_{\nu}- \partial _{\nu} A _{\mu}$,
the QED Lagrangian density in units such that $\hbar=c=1$ is
\begin{equation}\label{Eq01}
{\cal L}= -(1/4) F^{\mu \nu} F_{\mu \nu} +
 i\psi^{\dagger}\gamma^0\gamma^\mu D_\mu \psi .
\end{equation}
Coupling to the electromagnetic 4-potential $A_{\mu}$
occurs through the covariant derivative
\begin{equation}\label{Eq03}
D _{\mu} = \partial_{\mu}-ieA_{\mu} ,
\end{equation}
where -e is the renormalized electronic charge.
The notation used here defines covariant 4-vectors
\begin{eqnarray*}
x_{\mu} = ( t,- {\bf r}) , \;
\partial_{\mu} = (\partial/\partial t,\delv ) , \;
\\
A_{\mu} = (\phi , -{\bf A}) , \;
j_{\mu} = (\rho , -{\bf j}) .
\end{eqnarray*}
The metric tensor $g_{\mu\nu}$ is diagonal, with elements
$(1,-1,-1,-1)$.
Dirac matrices are represented in a form appropriate to a
2-component fermion theory, in which chirality $\gamma^5$
is diagonal for mass-zero fermions,
\begin{eqnarray}\label{Eq04}
\gamma^{\mu} =\left(\begin{array}{cc}
     0&I\\
     I&0 \end{array}\right),
\left(\begin{array}{cc}
     0& \sigv\\
  -\sigv&0 \end{array}\right);\hspace*{0.1in}
\gamma^5 =\left(\begin{array}{cc} -I&0\\ 0&I \end{array}\right).
\end{eqnarray}
\par Classical fields $\psi_p,A^\mu_k$ that satisfy homogeneous field
equations define quantum field operators
$ A^\mu(x)=\sum_k A^\mu_k(x) a_k$ and $\psi(x)=\sum_p\psi_p(x)\eta_p$.
Operators $a_k,\eta_p$ and fermion mass parameter $m_0$ define vacuum 
state $\rvac$ such that $a_k\rvac=0$ and $\eta_p\rvac=0$, where 
$\eta^\dag_p=\eta^\dag_a(\epsilon_a>0),
 \eta^\dag_p=\eta_i(\epsilon_i<0)$.
$A^\mu_k(x)=w^\mu_k e^{-ik.x}$ is the free photon field for 4-momentum
$k$. Dirac free-wave functions $\psi_p(x)$ are defined for
$\epsilon_p= p_0>0$ by
$\psi_p(x)=V^{-\half}e^{-ip.x}u_p, (\pstr-m_0)u_p=0$.
Reversing all components of 4-momentum for $\epsilon_p=-p_0<0$,
$\psi_{-p}(x)=V^{-\half}e^{ip.x}v_p, (\pstr+m_0)v_p=0$.
\par The standard model extends this formalism to include SU(2) gauge
fields.  Chiral projection operators, which force neutrino mass to
vanish, are included in Lagrangian interaction terms.
Because massless fermions have definite chirality,
$u_p\to u_{Lp}=\half(1-\gamma^5)u_p$ for sufficiently small mass,
and the chiral projection
operator $\half(1-\gamma^5)$ is redundant.
Because this operator is idempotent and commutes with
$\gamma^0\gamma_\mu$, for lepton-neutrino transitions $\ell\to\nu\ell$,
\begin{eqnarray}
 {\bar\psi}(\nu\ell)\gamma_\mu\psi(\ell)&\simeq&
 \psi^\dag(\nu\ell)\half(1-\gamma^5)\gamma^0\gamma_\mu\psi(\ell)
\nonumber\\
={\bar\psi}(\nu\ell L)\gamma_\mu\half(1-\gamma^5)\psi(\ell)&=&
 {\bar\psi}(\nu\ell L)\gamma_\mu\psi(\ell L) ,
\end{eqnarray}
if neutrino mass is sufficiently small.
If so, regardless of $m_\ell$, processes that depend on $\nu-W-\ell$
Feynman vertices, such as muon decay into charged leptons plus
neutrinos and V-A $\beta$-decay\cite{REN90,CAG98}, do not test the
standard-model insertion of such operators.  Neutrino self-interaction
is retained here by omitting them. 
Electroweak theory for quarks, involving hadronic structure,
is beyond the scope of the present analysis.
\par Although weak neutral current $Z^0_\mu$ vertices are not 
significant for lepton and neutrino mass, it is important to show that
the Lagrangian density assumed without chiral
projection can be consistently modified to give empirically correct 
neutral-current asymmetries\cite{CAG98,REN90}.  Neutral gauge current
$j^{(e)\mu}_{NC}={\bar\psi}_e\gamma^\mu(c_V-c_A\gamma^5)\psi_e$
is defined by ${\cal L}_{eZ}=
\frac{-e}{\sin2\theta}j^{(e)\mu}_{NC} Z^0_\mu$.
In the standard model\cite{CAG98},
$c_V=-\half+2\sin^2\theta, c_A=-\half$.
In Lagrangian density ${\cal L}_{\ell Z}$, without chiral
projection, $c_V=-\cos2\theta, c_A=0$.  
This can be converted to the SM form by the current density difference
$\Delta j^{\ell\mu}_{NC}=
  {\bar\psi}\gamma^\mu\Delta (c_V-c_A\gamma^5)\psi$
where $\Delta (c_V-c_A\gamma^5)=
 (-\half+2\sin^2\theta+\cos2\theta+\half\gamma^5)=
  \half(I+\gamma^5)$.
The incremental Lagrangian density is
$\Delta{\cal L}_{\ell Z}=
   \frac{-e}{\sin2\theta}\Delta j^{\ell\mu}_{NC} Z^0_\mu$.
This {\it ad hoc} correction has no undetermined parameters.
It preserves lepton-$W^\pm$ coupling for neutrino self-interaction.

\section{Algebraic formalism: mass as an eigenvalue}
\par The dynamical postulate implies field equations,
\begin{equation}\label{Eq05}
\partial^\mu F_{\mu\nu} = j_{\nu}
  = -e \psi^{\dagger}\gamma^0\gamma_{\nu}\psi, \;
i\gamma ^{\mu} D_{\mu}\psi = 0  .
\end{equation}
Because the current density satisfies $\partial_\mu j^\mu=0$, the
inhomogeneous Maxwell equation cannot change gauge condition
$\partial_\mu A^\mu_k=0$, if assumed for the photon fields.
Separating $A_\mu=A^{int}_\mu+A^{ext}_\mu$
into self-interaction and external
subfields, and using notation $\Astr(x)= \gamma^{\mu}A_{\mu}(x)$,
the fermion field equation
\begin{equation}\label{Dirmop}
\{i\gamma^\mu\partial_\mu+e\Astr^{ext}\}\psi=
  -e\Astr^{int}\psi={\hat m}\psi
\end{equation}
defines a self-interaction mass operator ${\hat m}=-e\Astr^{int}$. For
the quantized Maxwell field, the field amplitude operators have only
transition matrix elements.  Hence ${\hat m}$ is purely nondiagonal,
associated with virtual excitations of the radiation field.
The mass operator acts in a Fock space defined by products of creation
and annihilation operators.  There is no classical analog,
hence no valid classical model of elementary fermion mass.
\par What is proposed here is to rewrite Eq.(\ref{Dirmop})
as a renormalized Dirac equation and a mass-eigenvalue equation,
related by consistency condition $m_0=m$:
\begin{equation}\label{DirA}
\{i\gamma^\mu\partial_\mu+e\Astr^{ext}-m_0\}\psi=0, \;
 \{{\hat m}-m\}\psi =0.
\end{equation}
The second equation is solved in a Fock space defined by
mass parameter $m_0$, which is to be adjusted iteratively
to satisfy the consistency condition.
$m_0=0$ for bare fermions, but otherwise is a free parameter.
The computed eigenvalue can be identified with
$\delta m(m_0)$ in standard perturbation theory.  The consistency
condition $m_0=\delta m(m_0)$ forces the two equations to be equivalent
to the original field equation in renormalized Fock space.
\par Eqs.(\ref{DirA})
exhibit the algebraic structure implied by renormalization.
A canonical transformation of field operators and the vacuum state
diagonalizes the mass operator and determines a c-number mass.
This transformation mixes field components of positive and
negative energy, breaking chiral symmetry.  The resulting Dirac equation
combines chiral massless Weyl spinors into 4-component Dirac bispinors.
\par On a spacelike surface indexed by parameter $\tau$,
excitation operator $\chi^\dag(\tau)$ defines state
$\chi^\dag(\tau)\rvac$ and surface action integral
$W(\tau)=\lvac\int_\tau d^3x\chi{\cal L}\chi^\dag\rvac$.
Condition $\chi\rvac=0$, required for a stable pseudostate, may
imply a canonical transformation of the vacuum.
Defining $W=\int W(\tau)d\tau$ between nested spacelike surfaces,
coupled field equations for independent fields $\phi(x)$ are determined
by the variational condition
\\$
\delta W=\lvac\int d^4x\chi \delta \phi^\dag (
\frac{\delta{\cal L}}{\delta\phi^\dag}
-\partial_\mu\frac{\delta{\cal L}}{\delta\partial_\mu\phi^\dag }
)\chi^\dag\rvac=0.$
\par An algebraic theory is obtained by expanding
$\chi^\dag(\tau)=\sum_\lambda\chi^\dag_\lambda c_\lambda(\tau)$
as a sum of invariant excitation operators,
subject to $\chi(\tau)\rvac=0$.
If field Hamiltonian operator ${\hat H}(\tau)$ is defined on spacelike
surface $\tau$,  $\partial_\tau\chi^\dag=i[{\hat H},\chi^\dag]$
implies linear algebraic equations for the coefficients
$c_\lambda(\tau)$.  A field state is determined only if field and
algebraic equations are both satisfied.
Lagrange multiplier $m$, distinct from renormalization
parameter $m_0$, is required to enforce the normalization constraint
$\sum_\lambda|c_\lambda|^2=const$.
$\tau$-dependent phase factors are not determined, but can be
adjusted to remove diagonal elements of the algebraic equations.
This reduces them to the form of mass-eigenvalue equations.
\par Defining $W(\tau)$ on a spacelike surface clarifies several aspects
of the theory.  Past and future are well-defined.  A stable pseudostate
constructed on surface $\tau$ by superimposing global spacetime field
solutions requires consistency with past creation events (retarded
potentials) and with future annihilation events (advanced potentials).
This is built into Feynman propagators\cite{FEY49}.
\par For eigenvalue equations dominated by a particular excitation
operator $\chi^\dag_0$, it is convenient to normalize the state
eigenfunction so that $c_0=1$.  The basis set of elementary excitation
operators can be augmented systematically, adding terms which affect
higher orders in a perturbation expansion.  The algebraic equations have
a leading term $m=\sum_a(0|{\hat m}|a)c_a$, exact for nondiagonal 
${\hat m}$ if all nonzero matrix elements are included.
Using a formal solution for the coefficients $c_a$, $(a\neq 0)$,
\\$
 m=-\sum_{a,b}(0|{\hat m}|a)[{\hat m}-m]^{-1}_{ab}(b|{\hat m}|0).
$\\
This is a sum of terms that all vanish if $m\to\infty$.
The limit, an infinite sum of vanishing terms, may be finite, 
implying a cutoff inherent in the algebraic equations.  Rather than
explore the true limit of this expansion, derivations here will be
limited to the mass formula implied by the leading algebraic equation,
which acts as a mean value expression.  The gauge field equations
determine coefficients $c_a$ consistent with perturbation theory,
which would expand the inverse matrix here in powers of a coupling
constant.

\section{Mass of the electron}
\par The self-energy of a free electron can be evaluated to order $e^2$
using Feynman's rules\cite{FEY49,PAS95}.  The relevant Feynman diagram
describes virtual emission and reabsorption of a photon of 4-momentum
$k$ by a free electron.  The present analysis replaces perturbation
theory for this virtual process by approximate solution of eigenvalue 
equation $\{{\hat m}-m\}\psi=0$.  Renormalized electric charge is 
assumed.
\par If a solution of the inhomogeneous Maxwell equation is substituted
into the leading term of the mass-eigenvalue equation, it becomes
$m\psi_p(x)= m\int d^4y G_D(x,y)\psi_p(y)=
\int d^4yG_{{\hat m}D}(x,y)\psi_p(y),$
defining a modified propagator
\\$
G_{{\hat m}D}(x,y)=\sum_ke^2\gamma^\mu
\sqr^{-2}(x,y) \psi_{p-k}(x){\bar\psi}_{p-k}(y)\gamma_\mu.
$\\
Here $\sqr^{-2}(x,y)=G_A(x,y)$, the photon Green function of Feynman,
and $G_D(x,y)$ is the Feynman Dirac propagator.
The Fourier transform defines a convolution integral over 4-momentum,
$m(m_0)=\frac{ie^2}{(2\pi)^4}\int d^4k{\bar u}_p{\tilde G}^{\mu\nu}_A(k)
\gamma_\mu{\tilde G}_D(p-k)\gamma_\nu u_p$,
independent of $p$ due to Lorentz covariance.
This is the 2nd order Feynman self-energy\cite{FEY49}.
\par In a basis of massless chiral-projected (Weyl) spinors, the
convolution integral for $m(m_0)$ is purely nondiagonal.   Defining
$u_{pL}=\half(1-\gamma^5)u_p$ and $u_{pR}=\half(1+\gamma^5)u_p$,
then ${\bar u}_{pL}=u^\dag_p\half(1-\gamma^5)\gamma^0
 ={\bar u}_p\half(1+\gamma^5)$, so that spinor algebra implies
${\bar u}_{pL}\gamma_\mu\gamma_\nu u_{pL}=0$.  Similarly,
${\bar u}_{pL}\gamma_\mu\gamma_\nu u_{pR}\neq 0$.
Elementary Weyl spinors are combined into a Dirac spinor, as
required for finite mass, only if nondiagonal chiral matrix elements
are retained.  This precludes neutrino self-interaction in the standard
model.  
\par The integral for $m(m_0)$ diverges logarithmically.  Convergence
is forced if a covariant Feynman cutoff factor\cite{FEY48}
$C(k^2)=\int_0^\infty
 \frac{-\lambda^2}{k^2-\lambda^2}G(\lambda)d\lambda$,
where $\int_0^\infty G(\lambda)d\lambda=1$,
is inserted into the integrand.  If this is simplified to
$C(k^2)=\frac{-\Lambda^2}{k^2-\Lambda^2}$,
the integral converges to $m(m_0)=
\frac{3\alpha}{4\pi}m_0\left(\ln\frac{\Lambda^2}{m_0^2}+\half\right)$,
where $\alpha=e^2/4\pi$.
Consistency condition $m_0=m=m_e$ requires $\Lambda\simeq m_e e^{287}$.
This very large value may have cosmological significance, related to
a total mass or energy that sets an upper bound for intermediate virtual
photon momenta, or to a spatial microstructure such as that considered
in string theory.

\section{Neutrino mass} 
\par The self-interaction mass of the neutrino involves two distinct
weak-interaction processes.   The first, virtual excitation of an
electron/positron and a $W^\pm$ gauge boson, dominates the second,
virtual excitation of a neutrino and $Z^0$, because interaction mass 
is proportional to intermediate fermion mass.
SU(2) symmetry can be assumed for the intermediate $W^\pm$
process, eventually broken by the self-interaction 
masses of the $\nu,e$ doublet.
The SU(2) Lagrangian density\cite{YAM54} is
\\$
{\cal L}= -\frth {\bf W}^{\mu\nu}\cdot{\bf W}_{\mu\nu} +
 i{\bar\psi}\gamma^\mu D_\mu \psi,
$\\
where ${\bf W}_{\mu\nu}=\partial_\mu{\bf W}_\nu-\partial_\nu{\bf W}_\mu
 -g{\bf W}_\mu\times{\bf W}_\nu$.
Covariant derivative $D_\mu=\partial_\mu+i\frac{g}{2}\tW_\mu$
ensures SU(2) gauge invariance.  $\tauv$ is a 3-vector of
$2\times 2$ matrices identical to the Pauli matrices $\sigv$.
\par In the classical inhomogeneous SU(2) field equation, the fermion 
source current density is  
${\bf j}^\nu_W=\half g{\bar\psi}\gamma^\nu \tauv\psi$.
The nonabelian gauge field contributes an additional self-interaction
to a conserved total weak current density\cite{YAM54}.
It is assumed here that this self-interaction can be combined with
that induced by the Higgs boson to produce a field mass $M$ in an
effective inhomogeneous weak field equation
$\{\sqr^2+M^2\}{\bf W}^\mu={\bf j}^\mu_W$.
The Green function defined such that 
$
{\bf W}^\mu(x) =\int d^4y G_W^{\mu\nu}(x,y) {\bf j}_{W\nu}(y)
$\\
has the Fourier transform 
${\tilde G}^{\mu\nu}_W(k)=g^{\mu\nu}/(-k^2+M^2)$.
\par The neutrino self-interaction mass operator is 
${\hat m}_\nu=\frac{g}{2}\gamma^\mu\tW^{int}_\mu$. 
If $\{{\hat m}_\nu-m_\nu\}\psi=0$ and $m_{\nu0}=m_\nu$,
the renormalized neutrino field equation is
\\$\{i\gamma^\mu\partial_\mu-\frac{g}{2}\gamma^\mu\tW^{ext}_\mu
-m_{\nu0}\}\psi=0$.\\
In analogy to the electron self-interaction, the mass
eigenvalue can be estimated by the second-order Feynman 
convolution integral
\\$
m_\nu=\frac{ig^2/4}{(2\pi)^4} \int d^4k{\bar u}_p
{\tilde G}^{\mu\nu}_W(k)\gamma_\mu{\tilde G}_D(p-k)\gamma_\nu u_p.
$\\
Feynman propagator $G_D$ refers to an electron or positron, with
mass $m_e$, accompanied by gauge field $W^\pm$, 
restricted to ($t_y<t_x,t_y>t_x$), respectively.
An intermediate neutrino plus $Z^0$ implies an integral proportional
to neutrino mass, which can be neglected.
Spinors ${\bar u}_p\cdots u_p$ project onto the renormalized neutrino 
state.  Neglecting the neutrino mass parameter, $\pstr\simeq0$ 
simplifies the integral. The gauge field requires a Klein-Gordon 
propagator for $M=M_W$.  The Feynman 4-momentum cutoff for convergent 
integrals replaces the photon factor
\\$
C(k^2)=1-\frac{k^2}{k^2-\Lambda^2}=
 \frac{-\Lambda^2}{k^2-\Lambda^2}$ by\\
$C_M(k^2)=1-\frac{k^2-M^2}{k^2-M^2-\Lambda^2}=
 \frac{-\Lambda^2}{k^2-M^2-\Lambda^2}.
$
\par With these assumptions, using standard $\gamma$-matrix algebra,
the integral to be evaluated is 
\\$m_\nu(m_e)= \frac{ig^2}{4}\int\frac{d^4k}{(2\pi)^4}
 \frac{\Lambda^2}{k^2-M^2-\Lambda^2} \frac{1}{k^2-M^2}
{\bar u}_p\frac{2\kstr+4m_e}{k^2-2p\cdot k-m_e^2}u_p.$\\
The denominators here can be combined using \\
$\frac{1}{abc}=\int_0^1dy\int_0^1dx\frac{2y}{(axy+b(1-x)y+c(1-y))^3}.$\\
Setting $k'=k-p(1-y)$ and using $\pstr u_p\simeq 0$ removes 
$p\cdot k$ from the combined denominator and replaces $\kstr$ by 
$\pstr(1-y)\simeq 0$ in the numerator of the integral for $m_\nu$.    
Its final factor is effectively replaced by $4m_e/(k^2-m_e^2)$.  
Then $m_\nu(m_e)= 
ig^2m_e\Lambda^2\int_0^1 2ydy\int_0^1dx \; I_3$,
where 
\\$
I_3=\int\frac{d^4k}{(2\pi)^4}\frac{1}{(k^2-L)^3}= 
\int\frac{{\bf k}^2d|{\bf k}|}{2\pi^2} 
\int\frac{dk_0}{2\pi}\frac{1}{(k_0^2-({\bf k}^2+L))^3},
$\\ 
and $L=(M^2+\Lambda^2)xy+M^2(1-x)y+m_e^2(1-y)$.
The integrand has poles at $k_0=\pm\sqrt{{\bf k}^2+L}$.
Expanding $k_0^2={\bf k}^2+L+\zeta\epsilon+\epsilon^2$, where
$\zeta=2\sqrt{{\bf k}^2+L}$, a contour integral enclosing the positive 
pole (displaced below the real axis) implies 
\\$
\int\frac{dk_0}{2\pi}\frac{1}{(k_0^2-({\bf k}^2+L))^3}=
 -iRes_{\epsilon\to0}\frac{1}{\epsilon^3(\zeta+\epsilon)^3}=
\frac{6}{i\zeta^5}.
$\\
For $u^2={\bf k}^2+L$,
$\int_0^\infty\frac{{\bf k}^2d|{\bf k}|}{u^5}=\frac{1}{3L}$. \\
Hence $I_3=\frac{2^{-5}}{2i\pi^2}\frac{6}{3L}=\frac{1}{32i\pi^2L}$.
From this analysis, 
\\$ 
\frac{m_\nu}{m_e}=
\frac{g^2}{16\pi^2}\int_0^1dy
\ln\left(\frac{m_e^2(1-y)+(M^2+\Lambda^2)y}{m_e^2(1-y)+M^2y}\right).
$
\par Unlike the photon, $W^\pm$ is unstable.  Its energy width 
$\Gamma$ suppresses the neutrino mass integrand in coordinate space by
$e^{-\Gamma|t_x-t_y|}$, establishing a dynamical scale for Feynman 
cutoff factor $C_M(k^2)$.
$C_M(k^2)=\int_0^\infty\frac{-\lambda^2}{k^2-M^2-\lambda^2}
G(\lambda)d\lambda$, for a weight function normalized such that 
$\int_0^\infty G(\lambda)d\lambda=1$. 
If $G(\lambda)=
\sqrt{\frac{2}{\pi\Lambda^2}}e^{-\half\lambda^2/\Lambda^2}$,
then
$\int_0^\infty G(\lambda)\lambda^2d\lambda=\Lambda^2$.
If $\Lambda\ll M$,
$C_M(k^2)\simeq\frac{-\Lambda^2}{k^2-M^2-\Lambda^2}$,  
and $\Lambda^2\simeq\Gamma^2$ determines the cutoff factor. 
If $m_e\ll\Lambda\ll M$, $\frac{m_\nu}{m_e}\simeq
\frac{g^2}{16\pi^2}\ln(1+\frac{\Lambda^2}{M^2})\simeq
\frac{g^2}{16\pi^2}\frac{\Lambda^2}{M^2}$.  Using experimental
parameters\cite{CAG98} $g=$0.6524, $m_e=$0.511MeV, $M=M_W=$80.33GeV, 
and setting $\Lambda=\Gamma_W=$2.64GeV\cite{REN90},
$m_\nu=0.291\times 10^{-5}m_e=$1.49eV,
within current experimental limits\cite{NEU04}.
\par The algebraic theory considered here supports the speculation that
fermion generations correspond to excitation
operators in the sequence $\eta^\dag,\eta^\dag\eta\eta^\dag,
\eta^\dag\eta\eta^\dag\eta\eta^\dag$, acting on the renormalized vacuum.
For example, the muon might correspond to a $\ell{\bar\ell}$ pair
stabilized by an external electron, while the $\mu$-neutrino is such
a pair stabilized by an external e-neutrino.
Decay mechanisms would agree with well-established muon decay. 
Observed neutrino mixing implies very similar masses for the three
generations of neutrinos\cite{NEU04}.
Transition matrix elements coupled to gauge fields would be the same
for all generations, so that all neutrinos might have comparable 
self-interaction mass.
For charged leptons, the intermediate virtual state would be the
charged lepton itself plus a photon.  This depends on the still unknown
mechanism for 4-momentum cutoff, which might differ substantially for
different lepton generations.   

\section{The SU(2) scalar field } 
\par Weinberg-Salam electroweak theory includes a term of the form
\cite{REN90,CAG98}
\\$
\Delta{\cal L}=-V(\Phi^\dag\Phi)=
-\lambda(\Phi^\dag\Phi-\phi_0^2)^2
$\\ 
in the postulated Lagrangian density ${\cal L}_\Phi$
of the Higgs scalar $SU(2)$ doublet $\Phi=(\Phi_+,\Phi_0)$.  
$\Phi=(0,\phi_0)$ determines a stable neutral vacuum ground state.
Setting parameter $\lambda=\frac{w^2}{2\phi_0^2}>0$, \\ 
$\Delta{\cal L}=-\half w^2\phi_0^2+w^2\Phi^\dag\Phi
-\frac{w^2}{2\phi_0^2}(\Phi^\dag\Phi)^2$.
The imaginary mass term $w^2\Phi^\dag\Phi$ in $\Delta{\cal L}$
would by itself destabilize the vacuum state.  It must be counteracted
in any viable theory.  This is accomplished by the biquadratic term
$-\frac{w^2}{2\phi_0^2}(\Phi^\dag\Phi)^2$. 
If ${\cal L}_\Phi=   
 (\partial^\mu\Phi)^\dag\partial_\mu\Phi+\Delta{\cal L}$, 
omitting gauge coupling, the field equation is 
$
(\sqr^2-w^2)\Phi=-\frac{\Phi^\dag\Phi}{\phi_0^2}w^2\Phi.
$
Electroweak theory considers an exact particular solution $\Phi$ such
that $\sqr^2\Phi=0$ and $\Phi_+=0$, $\Phi_0=\phi_0$, constant in      
spacetime.
$\Phi=(0,\phi_0+H/\sqrt{2})$ defines Higgs field $H$.
If $m_H=\sqrt{2}w$, the field equation reduces to 
\\$
(\sqr^2+m_H^2)H=
-\frac{3w^2H^2}{\sqrt{2}\phi_0}-\frac{w^2H^3}{2\phi_0^2}.$
Covariant derivative
$D_\mu=\partial_\mu+i\frac{g_1}{2}B_\mu+i\frac{g_2}{2}\tW_\mu$ implies 
mass proportional to $\phi_0$ for each gauge field coupled to $\Phi$.
The particular linear combination of $B_\mu,W^0_\mu$ that decouples
from $\Phi_0$ defines $A_\mu$, massless if $\Phi_+=0$.
\par In gauge-invariant ${\cal L}_\Phi=(D^\mu\Phi)^\dag D_\mu \Phi$,
the covariant derivatives couple isospin doublet components of an SU(2) 
scalar field to the weak gauge fields.  
Implied self-interaction terms can replace a separately postulated
$\Delta{\cal L}$.  The quadratic form of derivatives provides   
an imaginary mass term, counteracted by virtual gauge field emission
and reabsorption.  This replaces parameters $w^2$ and $\lambda$ by 
interaction terms. The renormalized scalar field
equation is equivalent to that assumed in electroweak theory. 
\par In $U(1)\times SU(2)$ gauge theory, covariant derivative
$D_\mu=\partial_\mu+\half i(g_1yB_\mu+g_2\tW_\mu)$
acts on gauge-dependent fields.
For real gauge fields $U_\mu$ such that $\partial^\mu U_\mu=0$,
Lagrangian density ${\cal L}_\Phi=(D^\mu\Phi)^\dag D_\mu \Phi$ 
for scalar field $\Phi$ implies the field equation
\\$
(\sqr^2-{\hat w}^2)\Phi(y)=
-i(g_1yB_\mu+g_2\tW_\mu)\partial_\mu\Phi(y),
$\\ 
where
\\$
{\hat w}^2=\frth(g_1yB_\mu+g_2\tW_\mu)(g_1yB^\mu+g_2\tW^\mu).
$\\
Self-interaction due to virtual intermediate weak
field quanta arises from the weak current density 
$j_\mu=\frac{\partial{\cal L}_\Phi}{\partial U^\mu}$,
coupled to $U^\mu$.
$w^2$ can be replaced in $\Delta{\cal L}$  by 
an eigenvalue or mean value of ${\hat w}^2$ computed for the virtual 
gauge fields generated by this current density. 
For $j_\mu\sim\Phi^\dag\frac{i\partial_\mu\Phi}{\Phi}\Phi$,
the 2nd order self-interaction is operationally
equivalent to a term $-\lambda(\Phi^\dag\Phi)^2$ in $\Delta{\cal L}$.
Coupled only to the weak gauge field $Z^0_\mu$, $\Phi_0$ should have a
self-interaction mass comparable to the neutrino.

\section{Relation to gravitational theory}
\par The SU(2) scalar field provides a possible link between the
standard model and gravitational theory\cite{MAN06}.
Invariant action integral $I_a=\int d^4x \sqrt{-g} {\cal L}_a$
determines covariant energy-momentum tensor 
$T_a^{\mu\nu}=-\frac{2\delta I_a}{\sqrt{-g}\delta g_{\mu\nu}}$.
For gravitational ${\cal L}_g$,
$W_g^{\mu\nu}=\frac{\delta I_g}{\sqrt{-g}\delta g_{\mu\nu}}$
implies field equation 
\\$W_g^{\mu\nu}=\half\sum_a T_a^{\mu\nu}$.
In standard Einstein/Hilbert theory, for Newton constant $G_N$
and Ricci scalar $R=g^{\mu\nu}R_{\mu\nu}$,
${\cal L}_g=\frac{-R}{16\pi G_N}$ implies field equation 
\\$
R^{\mu\nu}-\half g^{\mu\nu}R=
G^{\mu\nu}=8\pi G_N\sum_a T_a^{\mu\nu}.$
\\
In conformal gravitational theory\cite{MAN06},
\\$
{\cal L}_g=-2\alpha_g (R^{\mu\nu}R_{\mu\nu}-\thrd R^2)=
-2\alpha_g({\cal L}_2-\thrd{\cal L}_1).
$\\ 
The field equation is $-2\alpha_g W^{\mu\nu}=
-2\alpha_g(W_2^{\mu\nu}-\thrd W_1^{\mu\nu})=\half\sum_a T_a^{\mu\nu}$.
For a complex scalar field,
${\cal L}_\Phi=(\partial_\mu\Phi)^\dag\partial^\mu\Phi+w^2\Phi^\dag\Phi
-\lambda(\Phi^\dag\Phi)^2$.
$I_\Phi$ is scale invariant if $R^{\mu\nu}=-3w^2g^{\mu\nu}$,
or  $w^2=\frac{-R}{12}$.
The standard model sets $\lambda=\frac{w^2}{2\phi_0^2}$ and
postulates a particular solution $\Phi=\phi_0$.
This determines a
global solution of the coupled equations such that
$T_\Phi^{\mu\nu}=
-\sxth\phi_0^2G^{\mu\nu}+g^{\mu\nu}\lambda\phi_0^4$\cite{MAN06}.
\par In a geometry (Robertson-Walker) such that $W_g^{\mu\nu}=0$,
averaged uniform matter produces $T_{kin}^{\mu\nu}$, and the field
equation reduces to $T_\Phi^{\mu\nu}+T_{kin}^{\mu\nu}=0$.  Defining
${\bar\Lambda}=6\lambda\phi_0^2=3w^2$ and $8\pi{\bar G}=6/\phi_0^2$,
the field equation becomes\cite{MAN06}
$G^{\mu\nu}-{\bar\Lambda}g^{\mu\nu}=8\pi{\bar G}T_{kin}^{\mu\nu}$.
This has the same form as the Einstein field equation.  Effective
coupling constant ${\bar G}$ and cosmological constant ${\bar\Lambda}$
are determined by the scalar field parameters $\phi_0$ and $w$.
In a universe with negligible curvature, the cosmological
(Friedmann) equation\cite{MIC99} for Hubble constant $H_0$ is
$H_0^2=\frac{8\pi G}{3}\rho_m+\frac{\Lambda}{3}$.
This defines empirical parameters $\Omega_m=\frac{8\pi G}{3H_0^2}\rho_m$
and $\Omega_\Lambda=\frac{\Lambda}{3H_0^2}$.
\par Conformal theory implies ${\bar\Omega}_\Lambda=
\frac{6\lambda\phi_0^2}{3H_0^2}=\frac{w^2}{H_0^2}$.
Without self-interaction, empirical values\cite{WIK07} 
$\Omega_\Lambda\simeq 0.732$ and $H_0\simeq 70.9 km\,s^{-1} Mpc^{-1}$, 
with $\phi_0=180GeV$, imply Higgs mass 
$m_H=\sqrt{2}w\simeq 10^{-33}eV$, if all quantum field action
integrals are scale-invariant (conformal). 
Thus in the standard model, which omits self-interaction, only an
extremely small $m_H$ is compatible with the empirical Hubble constant. 
The present argument indicates that virtual excitation of $Z^0_\mu$ 
could determine the otherwise undetermined parameter $\lambda$, and  
might supplement the bare Lagrangian term $w^2=-R/12$ by a much larger
induced self-interaction, still very small compared with the induced 
dynamical mass of charged fermions or the gauge bosons.

\section{Implications and conclusions} 
\par
If fermions are SU(2) doublets, regardless of chiral projection, this 
excludes direct Higgs-fermion interaction.  Fermion mass must arise from
self-interaction.  This is shown here to explain the finite but small
mass of neutrinos, while providing a rationale for the existence of
fermion generations distinguished only by mass.
Applied to a scalar (Higgs) boson in an SU(2) manifold, the present 
analysis implies that its self-interaction mass could
arise solely from weak interactions, and might be very small, analogous 
to neutrino mass.  Such small mass is contrary to lower bounds deduced 
from experimental data\cite{GHKD90}, assuming Yukawa coupling of 
the Higgs boson to fermions.  Since such Yukawa coupling 
is not consistent with the theoretical model used here, implications 
regarding small Higgs mass should be reexamined.
\par The author is indebted to Professor Michael Peskin
for comments and helpful criticism.

\end{document}